# Condensation into ground state in binary string models

Sergei Viznyuk

**Abstract**

The ensemble of binary strings defined via strong-interaction model exhibits enhanced condensation (collapse) into ground state below certain temperature. The non-interaction model shows gradual accumulation into ground state as $T \rightarrow 0$

Computer models based on principle of statistical ensemble may behave similar to real physical systems, without assumption of interaction potentials or assumption of other physical laws. As we show in this article, an ensemble of binary strings constructed with less internal degrees of freedom is sufficient for the model to exhibit behavior similar to Bose-Einstein condensation (BEC) in trapped dilute atomic gases [1] under low temperature. On other hand, a model used to construct the ensemble with no reduction in internal degrees of freedom only shows gradual accumulation into ground state. The presented models associate the ensemble of equal-length binary strings with one-dimensional gas of particles of certain mass, momentum, and energy [2].

We construct an ensemble of binary strings from source binary string **B**. We use two different models to associate source string **B** with $N$ strings in the ensemble. Each string in resultant ensemble is $M$ bits in length. In order to have macroscopic thermodynamic properties defined the ensemble has to be in equilibrium state. The equilibrium state of the ensemble is achieved when source string **B** has random distribution of bits with probability $p=k/M$, $k$ being the number of set bits in string **B**.

## *C*-model

The first model, which we shall call *C*-model, has been introduced previously [2]. In *C*-model, the ensemble is constructed using a computable scalar measure $C_n$, which is the *Hamming distance* [3] between string **B** and **B** itself shifted by $n$ bits:

$$C_n = \sum_{i=1}^{M} B_i \oplus B_{i+n} , \quad 0 \leq n < N , \quad 0 < N \leq M \quad (1)$$

, where $B_i$ is $i^{th}$ bit of a string **B**, consisting of $M$ bits with values 0 or 1, and $\oplus$ is logical XOR operator; $B_{i+n} = B_{i+n-M}$ for $i+n>M$; $N$ is the total number of strings in the ensemble. The product $B_i \oplus B_{i+n}$, $1 \leq i \leq M$ is a string $C_n$ of $M$ bits. The scalar value $C_n$ is the sum of set bits in string $C_n$. The value $C_n$ is taken as *the observable measure* of the ensemble, and $N$ as the number of observations. All strings $C_i$ having the same observable measure $C_i$ constitute *macrostate i* of the ensemble. The adjusted [2] binomical distribution gives the population number $N_i$ of $i^{th}$ macrostate in equilibrium:

$$N_i^{binomial}(C_i) = \frac{2 \times N}{K} \cdot \left( \frac{\left(\frac{M}{K}\right)!}{\left(\frac{C_i}{K}\right)! \left(\frac{M-C_i}{K}\right)!} \cdot \left(\frac{\overline{C}}{M}\right)^{\frac{C_i}{K}} \cdot \left(1 - \frac{\overline{C}}{M}\right)^{\frac{M-C_i}{K}} \right) \quad (2)$$

, where $K = 1 - \sqrt{\left|1 - 2 \cdot \dfrac{\overline{C}}{M}\right|}$ ;  $\overline{C} = \dfrac{1}{N}\sum_i N_i \cdot C_i = \dfrac{1}{M}\sum_{n=0}^{M-1} C_n = 2 \cdot M \cdot p \cdot (1-p)$

Distribution (2) for values of $p$ not too close to 0 or 1 is approximated by normal distribution

$$N_i^{normal}(C_i) = \frac{2 \times N}{\sqrt{2\pi \cdot M \cdot T}} \exp\left(-\frac{(C_i - \overline{C})^2}{2 \cdot M \cdot T}\right) = \frac{2 \times N}{\sqrt{2\pi \cdot M \cdot T}} \exp\left(-\frac{E_i}{T}\right) \quad (3)$$

, where temperature $T$ and energy levels $E_i$ have been defined previously [2] as:

$$T = K \cdot \frac{\overline{C}}{M} \cdot \left(1 - \frac{\overline{C}}{M}\right) = \left(1 - \sqrt{\left|1 - 2 \cdot \frac{\overline{C}}{M}\right|}\right) \cdot \frac{\overline{C}}{M} \cdot \left(1 - \frac{\overline{C}}{M}\right) \quad (4)$$

$$E_i = \frac{(C_i - \overline{C})^2}{2 \cdot M} \quad (5)$$

The factor $2\times$ in front of $N$ in (2) and (3) appears because the same observable value $C_i$ is realized with $k$, and $M-k$ set bits in source string $\mathbf{B}$. In physical terms, it means the energy levels corresponding to odd $C_i$ numbers are forbidden, and the population of energy levels is doubled for even $C_i$ numbers. From (2) the population of the ground state ($C_i = \overline{C}$) is

$$\frac{N_0(T)}{N} = \frac{2}{K} \cdot \left(\frac{\left(\dfrac{M}{K}\right)!}{\left(\dfrac{\overline{C}}{K}\right)!\left(\dfrac{M-\overline{C}}{K}\right)!} \cdot \left(\frac{\overline{C}}{M}\right)^{\frac{\overline{C}}{K}} \cdot \left(1 - \frac{\overline{C}}{M}\right)^{\frac{M-\overline{C}}{K}}\right) \quad (6)$$

, where $\overline{C}(T)$ and $K(T)$ are computable from (4) functions of temperature $T$ for a given $M$. The value provided by expression (6) can be shown to be equal to

$$\frac{N_0(T)}{N} = \frac{2}{\sqrt{2\pi \cdot M \cdot T}} \quad (7)$$

, for any $T>0$. From (7) we see that at temperature $T_c = \dfrac{2}{\pi \cdot M}$ , $N_0$ becomes equal to $N$, i.e. all particles are in the ground state. For temperatures below $T_c$ the ensemble no longer behaves as ensemble of particles but rather as one single-particle state [4], which is the characteristic feature of Bose-Einstein condensation (BEC) in physical systems.

## *B*-model

We also consider **B**-model, where the ensemble is built as a set of $N$ non-overlapping substrings $\mathbf{B_n}$ of a larger string $\mathbf{B}$. Each substring $\mathbf{B_n}$ is $M$ bits in length. Scalar value $B_n$ is the sum of set bits in string $\mathbf{B_n}$. The value $B_n$ is taken as the observable measure of the ensemble. The population number $N_i$ of $i^{th}$ macrostate in **B**-model is given by regular binomial distribution:

$$N_i^{binomial}(B_i) = \frac{N \cdot M!}{B_i!(M-B_i)!} \cdot \left(\frac{\overline{B}}{M}\right)^{B_i} \cdot \left(1 - \frac{\overline{B}}{M}\right)^{M-B_i} = \frac{N \cdot M!}{B_i!(M-B_i)!} \cdot p^{B_i} \cdot (1-p)^{M-B_i} \quad (8)$$

The binomial distribution (8) for values of $p$ not too close to 0 or 1 is approximated by normal distribution

$$N_i^{normal}(B_i) = \frac{N}{\sqrt{2\pi \cdot M \cdot T}} \exp\left(-\frac{(B_i - \overline{B})^2}{2 \cdot M \cdot T}\right) = \frac{N}{\sqrt{2\pi \cdot M \cdot T}} \exp\left(-\frac{E_i}{T}\right) \quad (9)$$

, where temperature $T$ and energy levels $E_i$ are:

$$T = \frac{\overline{B}}{M} \cdot \left(1 - \frac{\overline{B}}{M}\right) = p \cdot (1-p) \quad (10)$$

$$E_i = \frac{(B_i - \overline{B})^2}{2 \cdot M} \quad (11)$$

From (8) the population of the ground state ($B_i = \overline{B}$) in **B**-model is

$$\frac{N_0(T)}{N} = \frac{M!}{\overline{B}!(M-\overline{B})!} \cdot \left(\frac{\overline{B}}{M}\right)^{\overline{B}} \cdot \left(1 - \frac{\overline{B}}{M}\right)^{M-\overline{B}} \quad (12)$$

, where $\overline{B}(T)$ is computable from (10) function of temperature $T$ for a given $M$.

## Comparison of **B**-model and **C**-model

Figure 1 shows the population of ground state in **B**-model (solid lines) and **C**-model (dash lines) as a function of temperature, for different string lengths $M$. The length of the strings $M$ in ensemble plays the role of mass of the particles in one-dimensional gas [2]. The graphs for **B**-model have been calculated using (12). Dash lines for **C**-model have been calculated using (6).

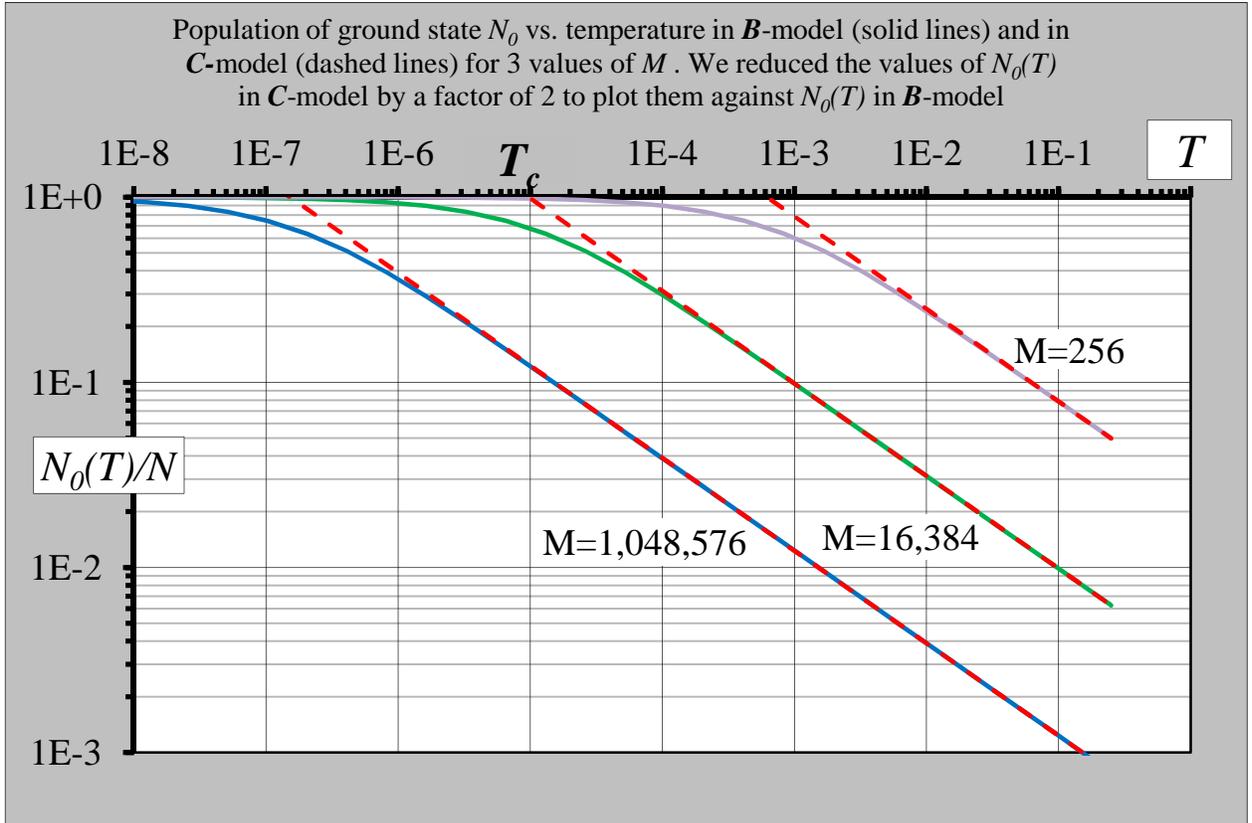

Figure 1.

The graphs show the ground state population in **B**-model gradually approaches $N$ as $T \rightarrow 0$, which is the expected behavior for an ideal gas of finite number of particles [4]. As temperature decreases, the population of ground state in **C**-model diverges from that in **B**-model. At some finite temperature $T_c > 0$ all particles in **C**-model condense into the ground state.

Figures 2 and 3 show what could be considered as experimental results. Several ensembles of binary strings have been prepared [5] for different temperature values. Figure 2 shows **C**-model distribution function in momentum space under 3 different temperature values. Markers on the graphs represent the "experimental" results, and solid lines represent the ajusted binomial distribution (2). The graph shows that under temperature $T=6.3E-5$, 77% of all particles are in zero-momentum state, with only 3 non-ground states showing as being populated with the rest 23% of the particles. This result is in agreement with (7). Also according to (7) for $M=16,384$ and temperature below $T_c = \dfrac{2}{\pi \cdot M} = 3.9E-5$ all particles collapse into the ground state.

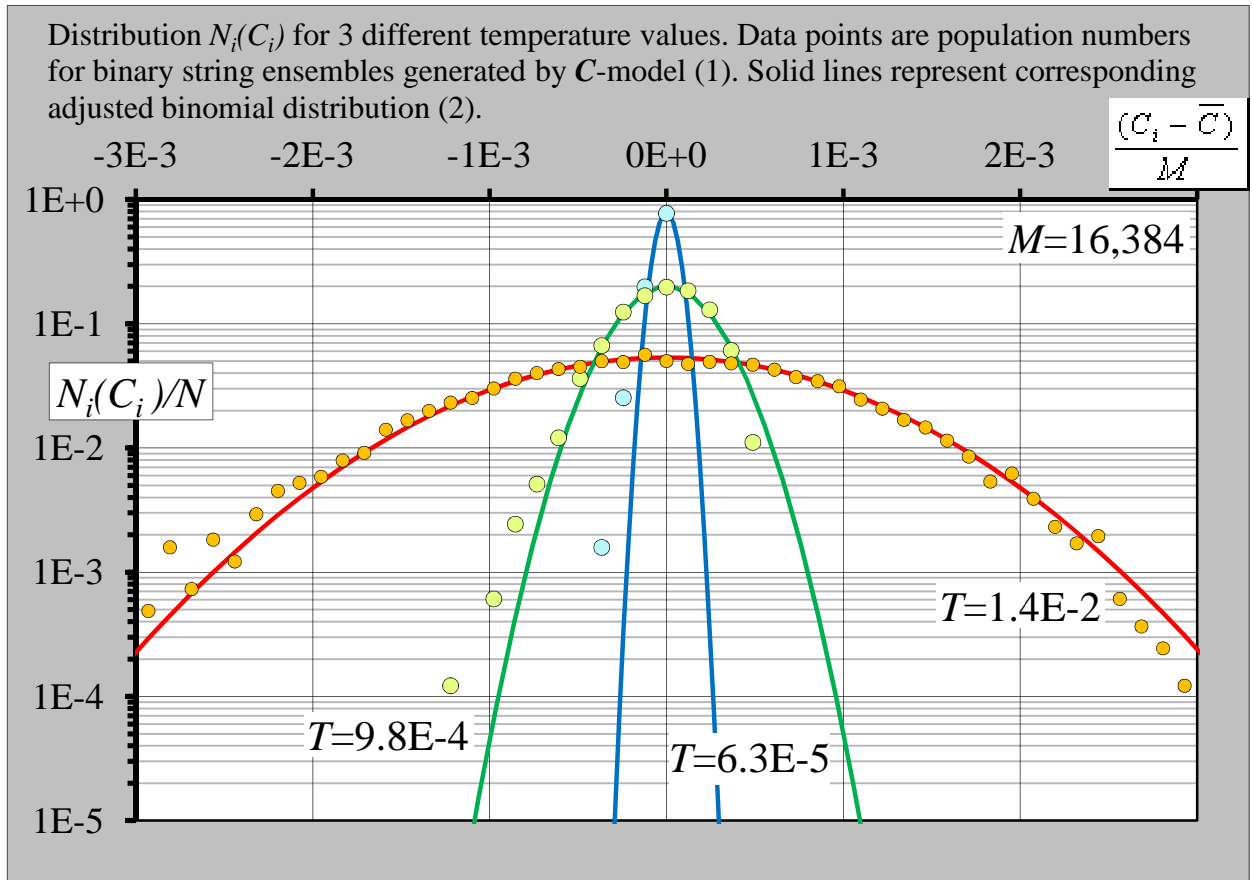

Figure 2

In contrast to $N_i(C_i)$ distribution in **C**-model, the $N_i(B_i)$ graphs in **B**-model on Figure 3 do not exhibit collapse into the ground state. Rather, as expected for an ideal gas consisting of finite number of particles [4], the population of the zero-momentum state gradually approaches total number of particles as $T \rightarrow 0$.

The ensembles in *C*-model and *B*-model consist of *N* binary strings of *M* bits in length. The ensembles considered to be in thermal equilibrium because of random distribution of set bits in source string. The probability of set bits in resultant ensemble strings is determined by temperature *T* according to (4) for *C*-model, and according to (10) for *B*-model. They key difference between ensemble in *C*-model and ensemble in *B*-model is that the strings in *B*-model ensemble are independent, while there is a strong correlation between strings in *C*-model by virtue of how ensemble in *C*-model is constructed using (1). We view such correlation as evidence of strong interaction between "particles" in *C*-model. Therefore, we consider *C*-model as strong-interaction model, and *B*-model as non-interaction model.

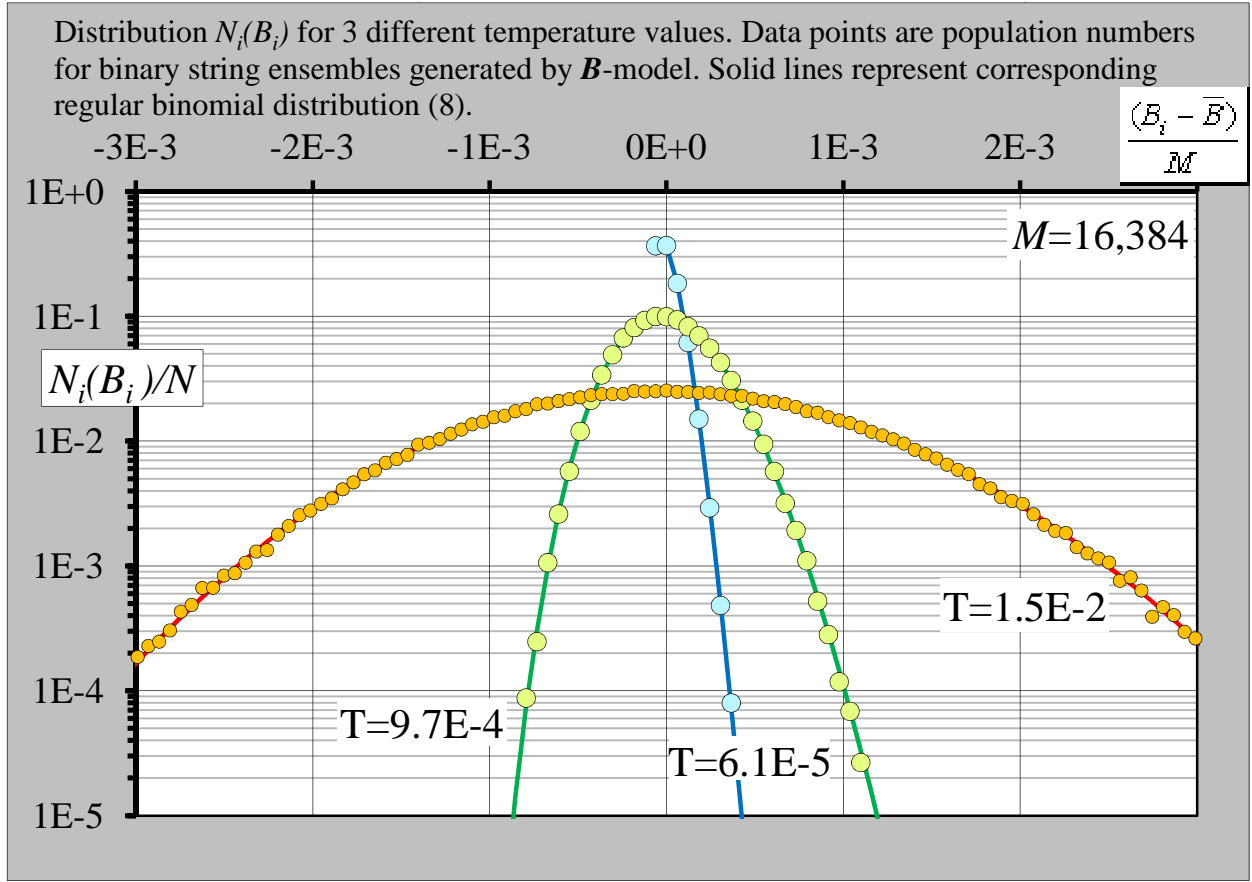

Figure 3

Strong interaction between particles in an ensemble leads to reduction in internal degrees of freedom, i.e. reduction in number of observable macrostates. Reduced number of allowed macrostates results in enhanced accumulation of particles into the ground state, as temperature decreases.

We have demonstrated by example of simple one-dimensional models that the interaction between particles can result in enhanced condensation (collapse) into the ground state in ensembles of finite number of particles under low temperatures. This result may relate to experimental observation of almost pure Bose-Einstein condensates in dilute atomic gases of finite number of particles confined in a magnetic trap [1,4].